\documentclass[aps,pra,twocolumn,showpacs,groupedaddress,floatfix]{revtex4}
\usepackage{amsfonts}
\usepackage{amsmath}
\usepackage{amssymb}
\usepackage{epsfig}
\usepackage{graphicx}
\usepackage{bm}
\usepackage{subfigure}
\usepackage{float}
\usepackage{color}

\begin{document}
\title{Precise quantum control via unsharp measurements and feedback operations}
\date{\today}
\author{Du Ran$^{1,2}$}
\author{Ye-Hong Chen$^{1,2}$}
\author{Zhi-Cheng Shi$^{1,2}$\footnote{szc2014@yeah.net}}
\author{Zhen-Biao Yang$^{1,2}$}
\author{Jie Song$^3$}
\author{Yan Xia$^{1,2}$\footnote{xia-208@163.com}}
\affiliation{$^1$Department of Physics, Fuzhou University, Fuzhou 350116, China\\$^{2}$Fujian Key Laboratory of Quantum Information and Quantum Optics (Fuzhou University), Fuzhou 350116, China\\$^3$Department of Physics, Harbin Institute of Technology, Harbin 150001, China}

%%%%%%%
\begin{abstract}
In this paper, we propose a scheme to eliminate the influence of noises on system dynamics, by means of a sequential unsharp measurements and unitary feedback operations.
The unsharp measurements are carried out periodically during system evolution, while the feedback operations are well designed based on the eigenstates of the density matrices of the exact (noiseless) dynamical states and its corresponding post-measurement states.
For illustrative examples, we show that the dynamical trajectory errors caused by both static and non-static noises are successfully eliminated in typical two-level and multi-level systems, i.e., the high-fidelity quantum dynamics can be maintained.
Furthermore, we discuss the influence of noise strength and measurement strength on the degree of precise quantum control.
Crucially, the measurement-feedback scheme is quite universal in that it can be applied to precise quantum control for any dimension systems.
Thus, it naturally finds extensive applications in quantum information processing.
\end{abstract}
\maketitle
%%%%%%%
\section{Introduction}

Precise control of quantum systems plays an important role in quantum science and technology.
How to effectively maintain high-fidelity manipulations thus becomes particularly critical, because it is the central requirement in several fields, such as fault-tolerant quantum computing \cite{Cha2000,MAN2000,AGa2002}.
Unfortunately, the fact that the interested system inevitably encounters various types of background noises always hinders the further development of quantum-enabled technologies \cite{WHZ2003,JQY2005,XWa2010,JMa2012}.

One of the biggest challenges for precise quantum control before \cite{ASm2013} is the lack of efficient, validated approaches in the presence of noises.
To overcome the influence of noises on system dynamics, many methods have been proposed recently, such as the composite pulse sequences \cite{TIc2011,TIc2013,Chin2014,XWa2012,JPKe2013,XWa2014,XWan2014}, optimal control \cite{GGo2008,RLK2013,BEA2015}, the invariant-based inverse engineering \cite{ARu2012,XJL2013}, filter-transfer-function \cite{TJGr2013,ASoa2014,GAPa2014}, adiabatic passages \cite{BoG2016}, single-shot shaped pulse \cite{DDa2013}, and so on.
In most of these methods, noise is assumed to be static, i.e., time independent during system evolution.
While for a general realistic situation, the interested system may suffer from stochastic time-varying  noise \cite{Chi2017}, $1/f^{\alpha}$ noise \cite{XCY2016}, or random classical noise \cite{JAu2012}, so that several methods would be invalid.
As a result, there is often badly in need of a simple and versatile protocol to eliminate the influence of all kinds of noises in different quantum systems.

On the other hand, unsharp measurement, a special positive operator-valued measure (POVM), has been widely used in quantum system control \cite{JAu2002,ZXM2012,SAs2010,KWa2014,MKP2014,WLM2018}, including tracking Rabi oscillations \cite{JAud2002,JAu2001,JAu2007,HBa2017}, investigating quantum process tomography \cite{HBa2015}, estimating full quantum state \cite{MHi2012}, monitoring wave function \cite{TKo2010}.
For unsharp measurement, one needs to design general measurement operators, which has been proposed in linear optical qubit system \cite{YOt2012}.
Recently, unsharp measurement with variable measurement-strength has been reported in experiment \cite{JTM2018}.
However, only using measurement in quantum system may bring about many restrictions in practice owing to the fact that the back-action effect of measurement is sometimes thought to be detrimental.
To sidestep such restrictions, one always combine measurement with feedback operations for quantum system control \cite{HUy2017,ACD2001,XXi2011,LMa2017,RVi2012,MSB2014,JJW2018,GGG2010}.
Indeed, it has been shown that the success probability for manipulating quantum system to target state  can be significantly improved by making use of feedback information  \cite{SSF2014},
and a conceptual guidance is developed to maintain quantum coherence by using unsharp measurements and unitary rotations in the presence of random classical noise fields \cite{JAu2012}.

In this work, we formulate a general measurement-feedback scheme to correct the dynamics of quantum system suffering from background noises.
We elaborately design the feedback operations with respect to the eigenstates of the density matrices of the exact (noiseless) dynamical states and its corresponding post-measurement states.
Such a designation method is general for any dimension scenario, and thus can be easily employed to different quantum systems.
For illustration purpose, we first focus on the static noises in typical qubit systems.
Then, we apply the scheme to eliminate the influence of time-varying noises in multi-level systems.
Numerical simulations show that the exact quantum system dynamics are achieved with high-accuracy via the measurement-feedback operations.
Notably, one prominent feature of the scheme is that there is no need to worry about the back-action effect of measurement, because such effect is fully compensated by the well-designed feedback operations.

The rest of this paper is organized as follows.
In Sec. \ref{systemmd}, we introduce the physical model and present the measurement-feedback operations.
In Sec. \ref{main}, we take some examples to show the performance of the scheme, including two-level and multi-level systems.
In Sec. \ref{discussions}, we discuss the influence of noise strength and measurement strength on the fidelity of system dynamical trajectory, then give a detailed analysis on experimental feasibility.
Conclusions are presented in Sec. \ref{conclusions}.

%Section 2
\section{Control model and measurement-feedback operations}\label{systemmd}
%\nonumber\\

Consider a general quantum system suffering from background noises, whose Hamiltonian is given by
\begin{eqnarray}\label{disH}
\hat{H}=\hat{H}_0+ \hat{H}_N,
\end{eqnarray}
where $\hat{H}_0$ is the undisturbed static Hamiltonian and $\hat{H}_N$ is the Hamiltonian of background noises.
In the absence of noises ($\hat{H}_N=0$), the system state exactly evolves as
\begin{eqnarray}\label{Eexa}
|\psi^E(t)\rangle= \hat{U}{(t)} |\psi(0)\rangle,
\end{eqnarray}
where $\hat{U}(t)=\exp({-i\hat{H_0}t})$, $|\psi(0)\rangle$ represents the initial state of system, and we set $\hbar=1$ hereafter.
In the following, the dynamical trajectory of exact (noiseless) system is labelled by DTES.
However, the real situation is that in presence of background noises the actual state of system
\begin{eqnarray}\label{Aexa}
|\psi^N(t)\rangle= \hat{\mathcal{U}}{(t)} |\psi(0)\rangle,
\end{eqnarray}
gradually deviates away from the exact dynamical state $|\psi^E(t)\rangle$, where $\hat{\mathcal{U}}{(t)} =\exp({-i\hat{H}t})$ if the background noises are unknown static.
When the background noises are time-varying, the evolution operator $\hat{\mathcal{U}}{(t)}$ in Eq. (\ref{Aexa}) should be replaced by
\begin{eqnarray}\label{TNU}
\hat{\mathcal{U}}(t)=\mathcal{T}\textrm{exp}\left\{ -i\int^{t}_{0}\hat{H}dt\right\},
\end{eqnarray}
where $\mathcal{T}$ is the time-ordering operator.
In the following, we label the dynamical trajectory of actual (noise) system by DTAS.

To eliminate the trajectory errors caused by the background noises (i.e., making DTAS converge to DTES), we employ a sequence of consecutive unsharp measurements and well-designed feedback operations.
The main control process is as below.
During system evolution process, we first perform a sequence of periodic unsharp measurements with period $\tau$. Note that the time of each measurement is assumed extremely short (e.g., adopt impulsive measurement approximation) so that it can be ignored during evolution.
For the $k$-th unsharp measurement with measurement result $n$ ($n=1, 2, \cdots, N$), it carried out on the actual state $|\psi^N(t_k)\rangle$ of the system, leading to the post-measurement state $|\psi^N_n(t_k)\rangle$:
\begin{eqnarray}\label{Measure}
|\psi^N_n(t_k)\rangle\equiv\frac{\hat{M}^{(k)}_n}{\sqrt{P^{(k)}_n}}|\psi^N(t_k)\rangle,
\end{eqnarray}
where the evolution time $t_k=k\tau$, $\hat{M}^{(k)}_n$ is the so-called Kraus operator corresponding to the measurement result $n$. $P^{(k)}_n=\langle \psi^N(t_k)|\hat{E}^{(k)}_n|\psi^N(t_k)\rangle$  is the detection probability of the measurement result $n$, with the effects $\hat{E}^{(k)}_n=\sqrt{(\hat{M}^{(k)}_n)^\dag (\hat{M}^{(k)}_n)}$, and the effects should satisfy $\sum_n\hat{E}^{(k)}_n=\hat{\mathbb{I}}$, where $\hat{\mathbb{I}}$ is the identity operator.
Generally speaking, after a sequence of unsharp measurements, the dynamical trajectory of post-measured-actual-system (DTPMAS) would deviate from the original one due to the back-action effect \cite{JAud2002,JAu2001,JAu2007,HBa2017,HBa2015}.
With such back-action effect, one can manipulate quantum states, including state initialization \cite{LRo2011} and entanglement generation \cite{WPf2013}.
However, it is unfavorable for eliminating the influence of noises on system dynamical trajectory, i.e., obtaining the DTES in the presence of  background noises.
One of the straightforward and effective strategy is to impose well-designed feedback operation after each measurement.
Here, the feedback operation is assumed to be instantaneous, i.e., the feedback process is accomplished without time consuming.
In addition, the feedback operation should satisfy the condition of restoring the corresponding post-measurement state of the exact state $|\psi^E(t_k)\rangle$ back into its pre-measurement state, i.e.,
\begin{eqnarray}\label{Un}
|\psi^E(t_k)\rangle=\hat{U}^{(k)}_n|\psi^E_n(t_k)\rangle,
\end{eqnarray}
where $|\psi^E_n(t_k)\rangle=\frac{\hat{M}^{(k)}_n}{\sqrt{P^{(k)}_n}}|\psi^E(t_k)\rangle$ is the post-measurement state after an unsharp measurement carried out on the exact state $|\psi^E(t_k)\rangle$.
Therefore, the DTES remains unchanged under the action of the sequence of measurement-feedback cycles.
In other words, the dynamical trajectory of the post-measured-exact-system (DTPMES) is the same with the DTES if the measurement-feedback cycles $\{\hat{U}^{(k)}_n\hat{M}^{(k)}_n\}$ are applied on the exact system.
By the sequence of measurement-feedback operations $\{\hat{U}^{(k)}_n\hat{M}^{(k)}_n\}$, the actual state of system evolves as
\begin{eqnarray}\label{NME}
|\psi^N_{M}(t_k)\rangle=\left(\prod^1_{k}\hat{U}^{(k)}_n\hat{M}^{(k)}_{n} \hat{\mathcal{U}}^{(k)}_{\tau}\right)|\psi(0)\rangle,
\end{eqnarray}
where $\hat{U}^{(k)}_{n}$ is the feedback operation, and $\hat{\mathcal{U}}^{(k)}_{\tau}$ is the evolution operator between two consecutive measurements. Note that $\hat{\mathcal{U}}^{(k)}_{\tau}=\exp(-i\hat{H}\tau)$ if the background noises are static, while
\begin{eqnarray}\label{TNU}
\hat{\mathcal{U}}^{(k)}_{\tau}=\mathcal{T}\textrm{exp}\left\{ -i\int^{t_k+\tau}_{t_k}\hat{H}dt\right\},
\end{eqnarray}
if the background noises are time-varying.
With these measurement-feedback operations, the DTPMAS would converge to the DTES eventually, indicating that we implement precise quantum control in the presence of background noises.

The physical mechanism why the DTPMAS can be driven to the DTES (i.e., the dynamical trajectory errors caused by background noises are cancelled by measurement-feedback operations) stems from following facts. When the same sequence of unsharp measurements carries out on two equal systems, the dynamical trajectory of the two systems would converge together \cite{HUy2017}.
Concretely, if we apply the same sequence of unsharp measurements on the actual system and the exact system, it would bring the DTPMAS and DTPMES converging together.
However, notice that the DTPMAS is still not converged to the DTES since the DTPMES and DTES are not equivalent, originating from the back-action effect of measurement.
To solve this intractable problem, we add the feedback operation after each measurement on the actual system,
where the feedback operation restores the post-measurement exact state back to its pre-measurement state (although there is no measurements carried out on the exact system).
Definitively, the DTPMES completely coincides with DTES under the sequence of measurement-feedback operations.
Thus the same measure-feedback operations would impose the DTPMAS converging to the DTES.
To gain a better understanding of the convergency precess, one can decompose the Kraus operator $\hat{M}^{(k)}_n$ into ``phase" and ``modulus" as complex number \cite{GKP1989}, i.e., $\hat{M}^{(k)}_n=\hat{\bar{U}}^{(k)}_n|\hat{M}^{(k)}_n|$, where $\hat{\bar{U}}^{(k)}_n$ is unitary.
Then, it is not hard to calculate that
\begin{eqnarray}\label{}
\hat{U}^{(k)}_n\hat{M}^{(k)}_n=\hat{U}^{(k)}_n\hat{\bar{U}}^{(k)}_n|\hat{M}^{(k)}_n| =\hat{\bar{U}}'^{(k)}_n|\hat{M}^{(k)}_n|=   \hat{M'}^{(k)}_n,
\end{eqnarray}
where $\hat{\bar{U}}'^{(k)}_n=\hat{U}^{(k)}_n\hat{\bar{U}}^{(k)}_n$ is an unitary operator.
As a reuslt, the measurement-feedback  operator $\hat{U}^{(k)}_n\hat{M}^{(k)}_n$ in fact can be regarded as Kraus operator $\hat{M'}^{(k)}_n$ as well.
In other words, the sequence of measurement-feedback operations $\{\hat{U}^{(k)}_n\hat{M}^{(k)}_n\}$ can be understood as a new sequence of unsharp measurements $\{\hat{M}'^{(k)}_n\}$.
With this new unsharp measurements applied on the actual system and the exact system, the DTPMAS would converge to the DTPMES.
As a result, the DTPMAS converges to the DTES, i.e., the dynamical trajectory errors caused by background noises are cancelled, by applying the measurement-feedback operations.

Finally, we elaborate on the designation of the feedback operations $\hat{U}^{(k)}_n$, which is based on the exact state $|\psi^E(t_k)\rangle$ and its post-measurement state $|\psi^E_n(t_k)\rangle$.
Here, we denote the corresponding density matrices of $|\psi^E(t_k)\rangle$ and $|\psi^E_n(t_k)\rangle$ by  $\rho^{E_k}=|\psi^{E_k}\rangle\langle\psi^{E_k}|$ and $\rho^{E_k}_n =|\psi^{E_k}_n\rangle\langle\psi^{E_k}_n|$ respectively, where we ignore the label $t_k$ for the sake of brevity.
Without loss of generality, in the $N$-dimensional Hilbert space, we suppose the concrete form of wavefunction $|\psi^{E_k}\rangle=[c_1, c_2, \cdots, c_j, \cdots, c_N]^T$, where $[\cdot]^T$ represents the transposition of the argument, and the complex numbers $c_j$ satisfy normalization condition: $\sum^N_{j=1}|c_j|^2=1$.
First, we construct a set of basis $\{|\phi^{E_k}_j\rangle\}$ to satisfy equation $\rho^{E_k}|\phi^{E_k}_j\rangle=\lambda^{E_k}_j|\phi^{E_k}_j\rangle$, whose expressions read
\begin{eqnarray}\label{}
 \begin{tabular}{c}
   $|\phi^{E_k}_1\rangle=|\psi^{E_k}\rangle=[c_1, c_2, \cdots, c_j, \cdots, c_N]^T,$ \\
   $|\phi^{E_k}_2\rangle=[-c^*_2, c^*_1, 0_3, \cdots, 0_j, \cdots, 0_N]^T$,  \ \ \ \ \\
   $|\phi^{E_k}_3\rangle=[-c^*_3, 0_2, c^*_1, \cdots, 0_j, \cdots, 0_N]^T$,  \ \ \ \ \\
   $\vdots$   \\
   $|\phi^{E_k}_j\rangle=[-c^*_j, 0_2, 0_3, \cdots, c^*_1, \cdots, 0_N]^T$,  \ \ \ \ \\
   $\vdots$   \\
   $|\phi^{E_k}_N\rangle=[-c^*_N, 0_2, 0_3, \cdots, 0_j, \cdots, c^*_1]^T$,  \ \ \ \
   \end{tabular}
\end{eqnarray}
where $0_j$ denotes that the $j$-th element is zero. The corresponding eigenvalues are $\lambda^{E_k}_1=1$ and $\lambda^{E_k}_j=0$ ($j=2, 3, \cdots, N$).
Clearly, the bases $|\phi^{E_k}_j\rangle$ are linearly independent with each other and completeness in the $N$-dimensional Hilbert space.
In fact, it composes of all eigenstates of density matrix $\rho^{E_k}$.
Then, by linear combination of  $|\phi^{E_k}_j\rangle$, we can define another set of general basis $\{|{\varrho}^{E_k}_m\rangle\}$, i.e., $|{\varrho}^{E_k}_m\rangle=\sum^N_{j=1}b_{j,m}|\phi^{E_k}_j\rangle$, where $b_{j,m}$ denote the normalized coefficients, $j, m=1, 2, \cdots, N$.
Analogously, we suppose $|\psi^{E_k}_n\rangle=[\tilde{c}_1, \tilde{c}_2, \cdots, \tilde{c}_{j'}, \cdots, \tilde{c}_N]^T$, where the complex numbers $\tilde{c}_{j'}$ satisfy $\sum^N_{{j'}=1}|\tilde{c}_{j'}|^2=1$.
Again, we construct a set of complete basis $\{|\tilde{\phi}^{E_k}_{j'}\rangle\}$ to satisfy equation $\rho^{E_k}_n|\tilde{\phi}^{E_k}_{j'}\rangle=\tilde{\lambda}^{E_k}_{j'}|\tilde{\phi}^{E_k}_{j'}\rangle$:
\begin{eqnarray}\label{}
 \begin{tabular}{c}
   $|\tilde{\phi}^{E_k}_1\rangle=|\psi^{E_k}_{n}\rangle=[\tilde{c}_1, \tilde{c}_2, \cdots, \tilde{c}_{j'}, \cdots, \tilde{c}_N]^T,$ \\
   $|\tilde{\phi}^{E_k}_2\rangle=[-\tilde{c}^*_2, \tilde{c}^*_1, 0_3, \cdots, 0_{j'}, \cdots, 0_N]^T$,  \ \ \ \ \\
   $|\tilde{\phi}^{E_k}_3\rangle=[-\tilde{c}^*_3, 0_2, \tilde{c}^*_1, \cdots, 0_{j'}, \cdots, 0_N]^T$,  \ \ \ \ \\
   $\vdots$   \\
   $|\tilde{\phi}^{E_k}_{j'}\rangle=[-\tilde{c}^*_{j'}, 0_2, 0_3, \cdots, \tilde{c}^*_1, \cdots, 0_N]^T$,  \ \ \ \ \\
   $\vdots$   \\
   $|\tilde{\phi}^{E_k}_N\rangle=[-\tilde{c}^*_N, 0_2, 0_3, \cdots, 0_{j'}, \cdots, \tilde{c}^*_1]^T$,  \ \ \ \
   \end{tabular}
\end{eqnarray}
with the corresponding eigenvalue $\tilde{\lambda}^{E_k}_1=1$ and $\tilde{\lambda}^{E_k}_{j'}=0$ (${j'}=2, 3, \cdots, N$).
Then, the set of general basis $\{|\tilde{\varrho}^{E_k}_{m'}\rangle\}$ can be $|\tilde{\varrho}^{E_k}_{m'}\rangle=\sum^N_{{j'}=1}\tilde{b}_{{j'},{m'}}|\tilde{\phi}^{E_k}_{j'}\rangle$, where $\tilde{b}_{{j'},{m'}}$ denote the normalized coefficients, $m', j'=1, 2, \cdots, N$.
After constructing the basis $\{|{\varrho}^{E_k}_m\rangle\}$ and $\{|\tilde{\varrho}^{E_k}_{m'}\rangle\}$, the explicit expression of the feedback operation $\hat{U}^{(k)}_n$ can be chosen as
\begin{eqnarray}\label{Feedb11}
\hat{U}^{(k)}_n=\sum_{m=1}^N\sum_{m'=1}^N|\varrho^{E_k}_m\rangle\langle\tilde{\varrho}^{E_k}_{m'}|.
\end{eqnarray}
Clearly, $\hat{U}^{(k)}_n$ is unitary (i.e., $(\hat{U}^{(k)}_n)^\dag \hat{U}^{(k)}_n=\hat{\mathbb{I}}$) and satisfies the condition of Eq. (\ref{Un}) since
\begin{eqnarray}\label{}
 \langle \psi^{E_k}|\psi^{E_k}\rangle &=& \langle \psi^{E_k}|\hat{U}^{(k)}_n|\psi^{E_k}_n\rangle \cr
&=& \sum^N_{m=1}\sum^N_{{m'}=1}\langle\psi^{E_k}|\varrho^{E_k}_m\rangle\langle\tilde{\varrho}^{E_k}_{m'}|\psi^{E_k}_n\rangle  \cr
&=& \sum^N_{\substack{m=1 \\ m'=1}}\sum^N_{\substack{j=1 \\ j'=1}}b_{j,m}\tilde{b}_{{j'},{m'}}\langle\psi^{E_k}|\phi^{E_k}_j\rangle\langle\tilde{\phi}^{E_k}_{j'}|\psi^{E_k}_n\rangle \cr
&=& 1.
\end{eqnarray}
Here we have used the orthogonality relations: $\langle\phi_{1}^{E_k}|\phi^{E_k}_j\rangle=\delta_{1j}$ and $\langle\tilde{\phi}^{E_k}_{j'}|\tilde{\phi}^{E_k}_{1}\rangle=\delta_{j'1}$.
It is worth mentioning that the simplest way to construct $\hat{U}^{(k)}_n$ is to choose $b_{j,m}=\delta_{j,m}$ and $\tilde{b}_{{j'},{m'}}=\delta_{j',m'}$, which lead to $|\varrho^{E_k}_m\rangle=|\phi^{E_k}_m\rangle$ and $|\tilde{\varrho}^{E_k}_m\rangle=|\tilde{\phi}^{E_k}_{m}\rangle$, respectively.
Then Eq. (\ref{Feedb11}) can be further simplified as
\begin{eqnarray}\label{Feedb}
\hat{U}^{(k)}_n=\sum_m^N|\phi^{E_k}_m\rangle\langle\tilde{\phi}^{E_k}_m|.
\end{eqnarray}
Note that in the construction process of feedback operation $\hat{U}^{(k)}_n$, there is no restrictions for the dimension $N$ of the system.
In other words, this method is completely general for any dimension systems.

\section{Examples}\label{main}

\subsection{Precise control in qubit system}

In this subsection, we apply the measurement-feedback scheme to eliminate the dynamical errors caused by static noises in qubit system, e.g., a spin-$\frac{1}{2}$ particle processing in a magnetic field.
In the absence of static noises, the Hamiltonian of the system is given by
\begin{eqnarray}\label{spsinH}
\hat{H}_{0}=\frac{\Omega_L}{2}\hat{{\bm{r}}}\cdot\hat{\bm{\sigma}},
\end{eqnarray}
where $\Omega_L$ is the Larmor frequency, $\hat{\bm{\sigma}}=(\sigma_x, \sigma_y, \sigma_z)$ is the Pauli matrices that generate rotation on $x, y, z$ axis, respectively.
The normalized vector $\hat{{\bm{r}}}=(r_x, r_y, r_z)$ represents the direction of the magnetic field.
Given an initial state, the spin system would exactly evolve as Eq. (\ref{Eexa}).
When existing static noises in the Larmor frequency and the direction of the magnetic field, the form of system Hamiltonian is denoted as
\begin{eqnarray}\label{}
\hat{H}=\frac{\Omega_L+ \Omega_\varepsilon}{2}\hat{{\bm{r}}'}\cdot\hat{\bm{\sigma}},
\end{eqnarray}
where $\Omega_\varepsilon$ quantifies the amount of static noise in Larmor frequency and $\hat{{\bm{r}}}'=(r'_x, r'_y, r'_z)$ denotes the direction of magnetic field in the presence of static noise.
Then, the actual spin system would evolve as Eq. (\ref{Aexa}).

\begin{figure}[htpb]
\centering
\scalebox{0.45}{\includegraphics{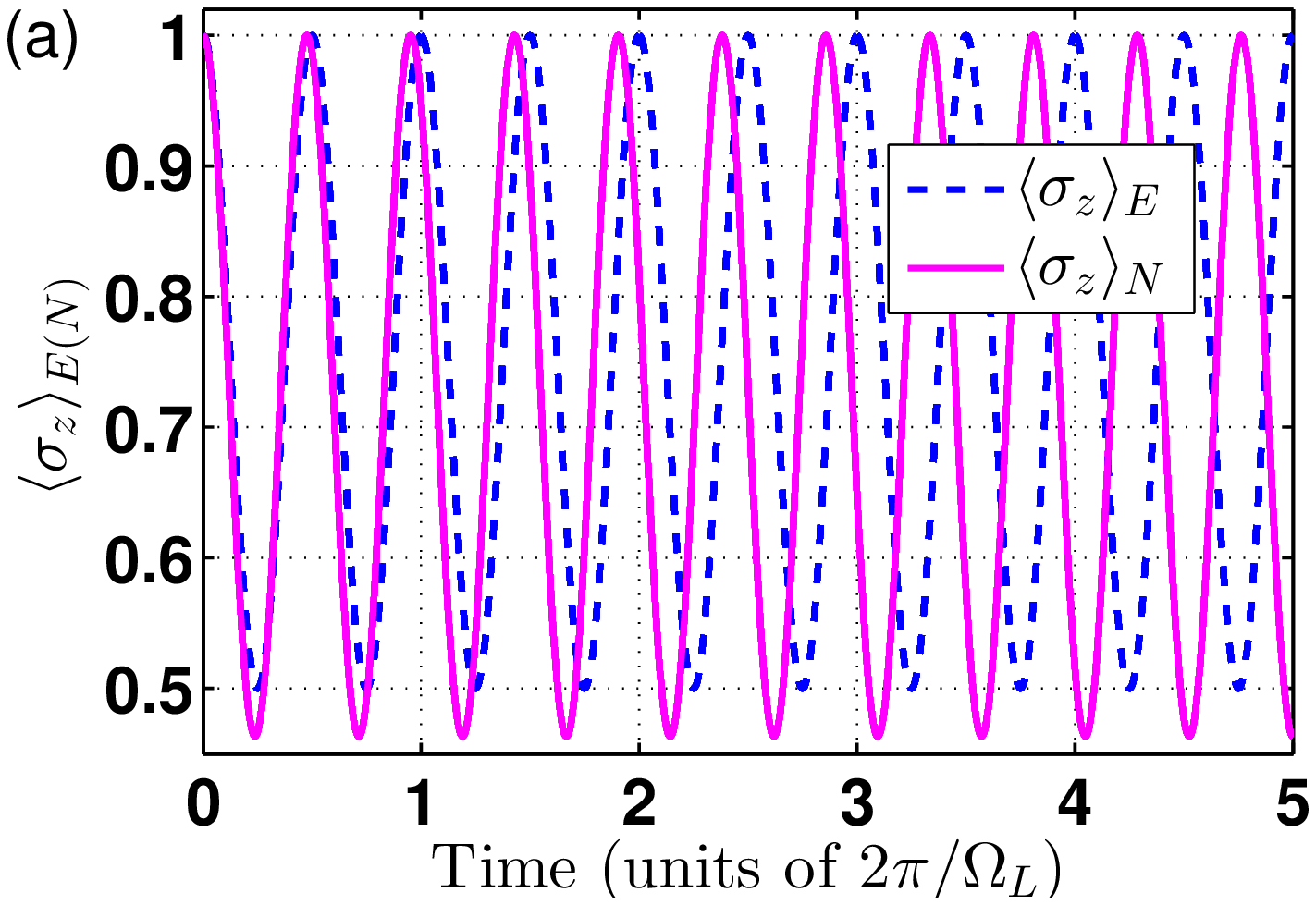}}
\scalebox{0.45}{\includegraphics{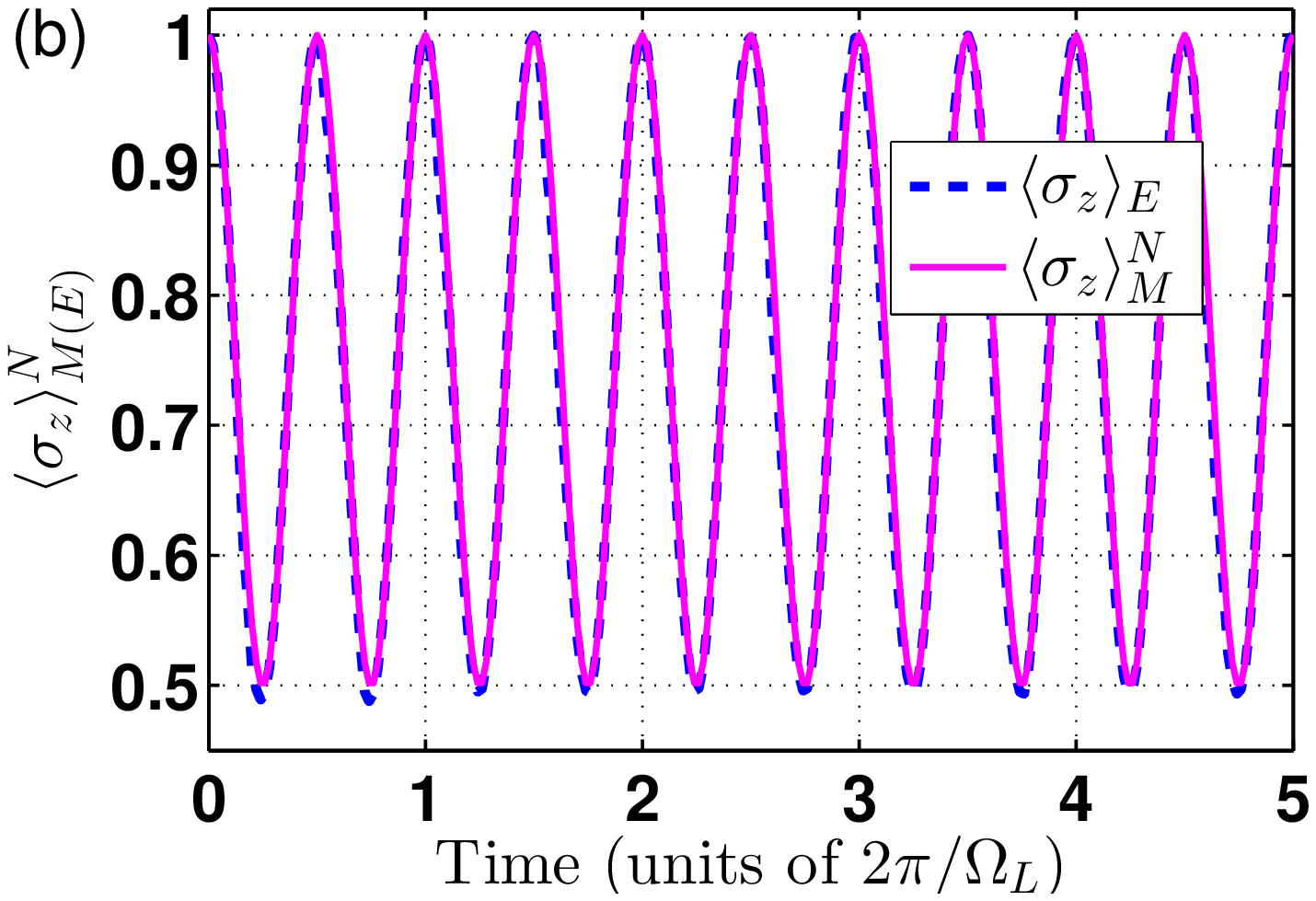}}
\caption{ Time evolution of the expectation values of $\sigma_z$ for the spin system in the absence of measurement-feedback operations (a), and in the presence of measurement-feedback operations (b). Parameters for the simulations are: $\Omega_\varepsilon=0.05\Omega_L$, $\theta=\pi/3$, $\phi=\pi/4$, $\theta'=\theta-\theta/50$, $\phi'=\phi-\phi/50$, $p_0=0.2$, $K=250$, and $\tau=2\pi/(50\Omega_L)$.}\label{FeildB}
\end{figure}

For the purpose of illustration, we now parameterize the noise in Larmor frequency as $\Omega_{\varepsilon}=0.05\Omega_L$, the original direction of the magnetic field as $\hat{\textbf{{r}}}=(\cos\theta\cos\phi, \cos\theta\sin\phi, \sin\theta)$, the noise-disturbed direction as $\hat{{\bm{r}}}'=(\cos\theta'\cos\phi', \cos\theta'\sin\phi', \sin\theta')$, and
the initial state as $|\psi(0)\rangle=|+\rangle$ with $|\pm\rangle$ being the eigenstates of $\sigma_z$.
%We exhibit the evolution of the system via the expectation value of $\sigma_z$.
In figure \ref{FeildB}(a), the blue dash line displays the exact evolution trajectory of the expectation value $\langle\sigma_z\rangle_E$ ($\equiv\langle\psi^E|\sigma_z|\psi^E\rangle$), while the magenta solid line displays the noise-disturbed evolution trajectory of the expectation value $\langle\sigma_z\rangle_N$ ($\equiv\langle\psi^N|\sigma_z|\psi^N\rangle$).
It is clear that not only the amplitude of the oscillation is changed, but also a phase shift of the oscillation frequency is induced by static noises, meaning that the static noises drive the dynamical trajectory of the system far away from the noiseless one.

The aim here is to adapt a sequence of unsharp measurements and feedback operations to impose the dynamical trajectory of noise-disturbed system approaching to the dynamical trajectory of noiseless one.
We would like to emphasis that although the dynamical trajectory of noiseless system can be monitored via unsharp measurements \cite{JAu2012}, there generally exists unavoidable phase kicks between post-measured system and original system due to the back-action effect of measurements \cite{JAu2007}.
With the help of feedback operations, the dynamical trajectory of noiseless system is achievable.
For the spin system, we perform unsharp measurements of the $\sigma_z$ observable with the Kraus operators given by \cite{HBa2015}
\begin{eqnarray}\label{MeOp}
\hat{M}^{(k)}_0=\sqrt{1-p_0}|-\rangle\langle -|+\sqrt{p_0}|+\rangle\langle +|,\cr
\hat{M}^{(k)}_1=\sqrt{p_0}|-\rangle\langle -|+\sqrt{1-p_0}|+\rangle\langle +|,
\end{eqnarray}
where $\sum_n(\hat{M}^{(k)}_n)^\dag\hat{M}^{(k)}_n=\hat{\mathbb{I}}$ ($n=0, 1$) and $0<p_0<0.5$.
The strength of a single measurement is quantified by $\Delta p=1-2p_0$, and $\Delta p\rightarrow0$ ($\Delta p\rightarrow1$) represents a fully weak (strong) measurement of observable.
Note that the strength of a sequence of measurements depends on not only $\Delta p$ but also the measurement frequency $1/\tau$.
The more frequent of the measurements are applied to the system, the more information we obtain \cite{JAu2012}.
We now choose $p_0=0.2$ and carry out a measurement-feedback operation every $\tau=T_L/50$, where $T_L=2\pi/\Omega_L$ is the oscillation period.
The related results are shown in Fig. \ref{FeildB}(b) by the evolution trajectory of expectation value  $\langle\sigma_z\rangle^{N}_{M}$ ($\equiv\langle\psi^N_M|\sigma_z|\psi^N_M\rangle$), where $|\psi^N_M\rangle$ represents the state after measurement-feedback operations, i.e., Eq. (\ref{NME}).
We can find that there is no phase shift phenomena between the noiseless evolution (blue dash line) and the evolution under measurement-feedback operations (magenta solid line), which indicates that the influence of static noises on the dynamical trajectory of qubit system is eliminated very well by measurement-feedback operations.

%%%%%%%
\subsection{Precise control in multi-level system}\label{Generaliz}

In the above investigation, we have shown that the measurement-feedback operations can eliminate the influence of static noises in the single two-level system.
In this subsection, we show the capability of the measurement-feedback scheme for the multi-level systems.
Here, we consider a more general case that the system suffers from time-varying noises, where the noise Hamiltonian is denoted by
\begin{eqnarray}\label{NoisH}
\hat{H}_N =\sum^L_{l=1}\lambda_l(t)\hat{\mathcal{H}}_l.
\end{eqnarray}
where $\lambda_l(t)$ represent the time-varying noise fields.
For simplicity, we assume the noises obey Gaussian distribution, i.e., $\lambda_l(t)=\frac{1}{\sqrt{2\pi}\sigma}\exp[-\frac{(t-\mu)^2}{2\sigma^2}]$, where $\mu$ and $\sigma$ represent the mean value and standard deviation of Gaussian distribution, respectively.
Under the influence of time-varying noises, it is no doubt that the dynamical trajectory of the system deviates away from the noiseless one.
For a single $N$-level system whose bare states are marked by $\{|d\rangle, d=1, 2, \cdots, N\}$, the Kraus operator $\hat{M}^{(k)}_n$ ($n=1, 2, \cdots, N$) can be constructed as
\begin{eqnarray}\label{Effects}
\hat{M}^{(k)}_n = \sum_{d=1}^N\sqrt{p_{d}}|d\rangle\langle d|,
\end{eqnarray}
where $p_d$ quantifies the strength of a single measurement and satisfy the relation $\sum^N_{d=1}p_d=1$ ($0<p_d<1$).
The single measurement strength is stronger when the value of $|p_d-1/N|$ is larger.

\begin{figure}[h]
\centering
\scalebox{1}{\includegraphics{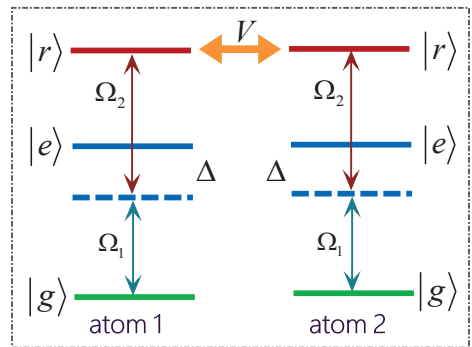}}
\caption{ Schematic representation of two trapped Rydberg atoms. $|r\rangle$ is the Rydberg state, while $|g\rangle$ and $|e\rangle$ are the ground and excited state, respectively. $V$ denotes the Rydberg-Rydberg-interaction strength. The ground states $|g\rangle$ and Rydberg state $|r\rangle$ are dispersively coupled to the excited state $|e\rangle$ with Rabi frequencies $\Omega_1$ and $\Omega_2$, respectively. $\Delta$ represents the corresponding detuning parameter.}\label{Rydberatom}
\end{figure}

Next, we demonstrate the measurement-feedback scheme by considering two identical Rydberg atoms coupled by laser fields. As shown in Fig. \ref{Rydberatom}, each atom has ground state $|g\rangle$, excited state $|e\rangle$ and Rydberg state $|r\rangle$.
The ground state $|g\rangle$ is dispersively coupled to the excited state $|e\rangle$ by a laser field with Rabi frequency $\Omega_1$ and detuning $\Delta$.
The excited state $|e\rangle$ can be pumped into the Rydberg state $|r\rangle$ by a laser field with Rabi frequency $\Omega_2$ and detuning $-\Delta$.
The Rydberg-Rydberg-interaction strength is $V$.
In the interaction picture, the Hamiltonian of the system reads
\begin{eqnarray}\label{undistH}
\hat{H}_I &=& [(\Omega_1|e\rangle_1\langle g| + \Omega_2|r\rangle_1\langle e|)\otimes\hat{\mathcal{I}}_2 \cr &+&\hat{\mathcal{I}}_1\otimes(\Omega_1|e\rangle_2\langle g| + \Omega_2|r\rangle_2\langle e|) + H.c.] \cr
     &+& \Delta(|e\rangle_1\langle e|\otimes\hat{\mathcal{I}}_2+\hat{\mathcal{I}}_1\otimes|e\rangle_2\langle e|) +V|rr\rangle\langle rr|,\nonumber\\
\end{eqnarray}
where $|mn\rangle$ is the abbreviation of $|m\rangle_1|n\rangle_2$ and $\hat{\mathcal{I}}_j$ ($j=1, 2$) denotes the $3\times3$ identity matrix.

\begin{figure}[]
\centering
\scalebox{0.45}{\includegraphics{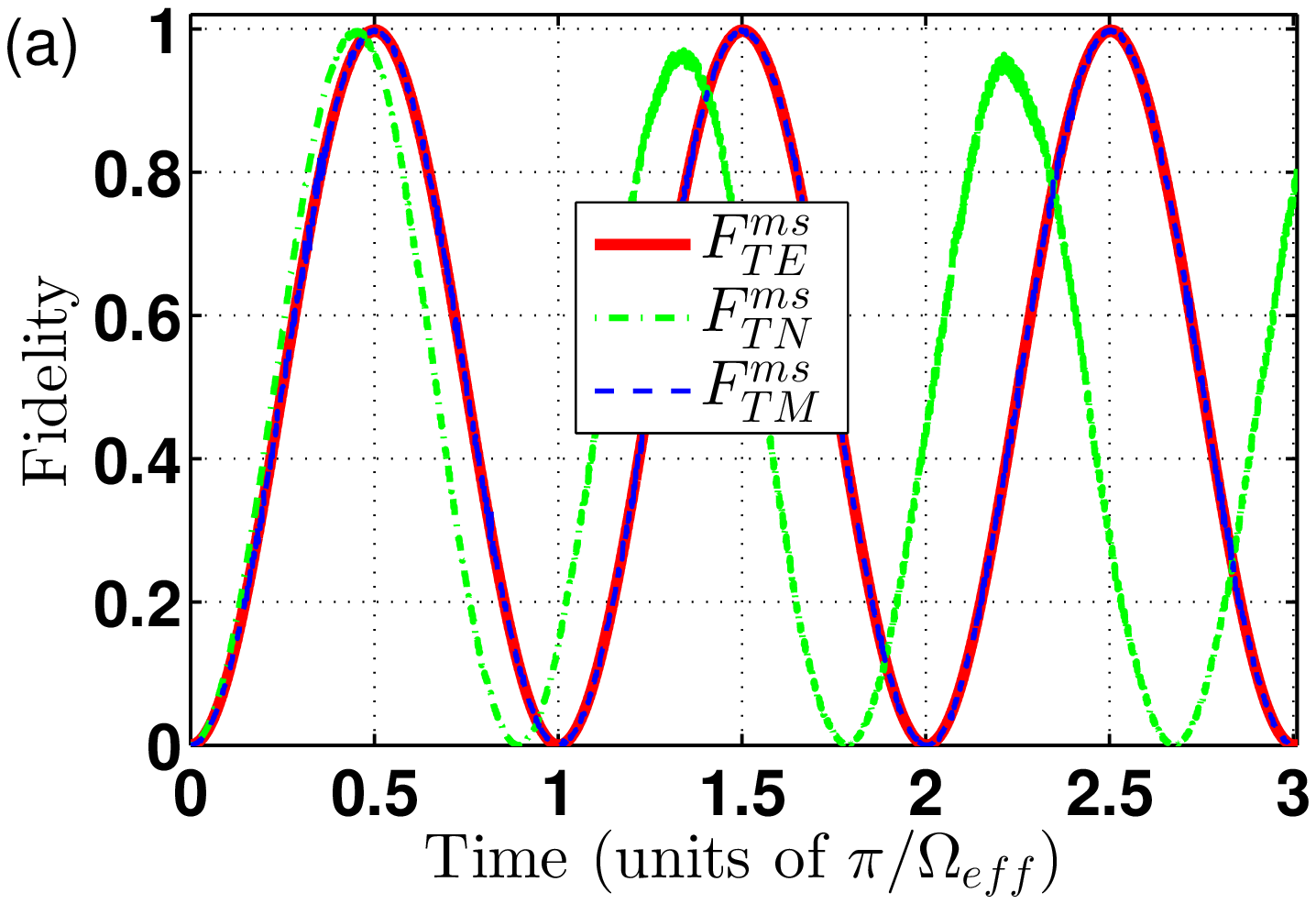}}
\scalebox{0.45}{\includegraphics{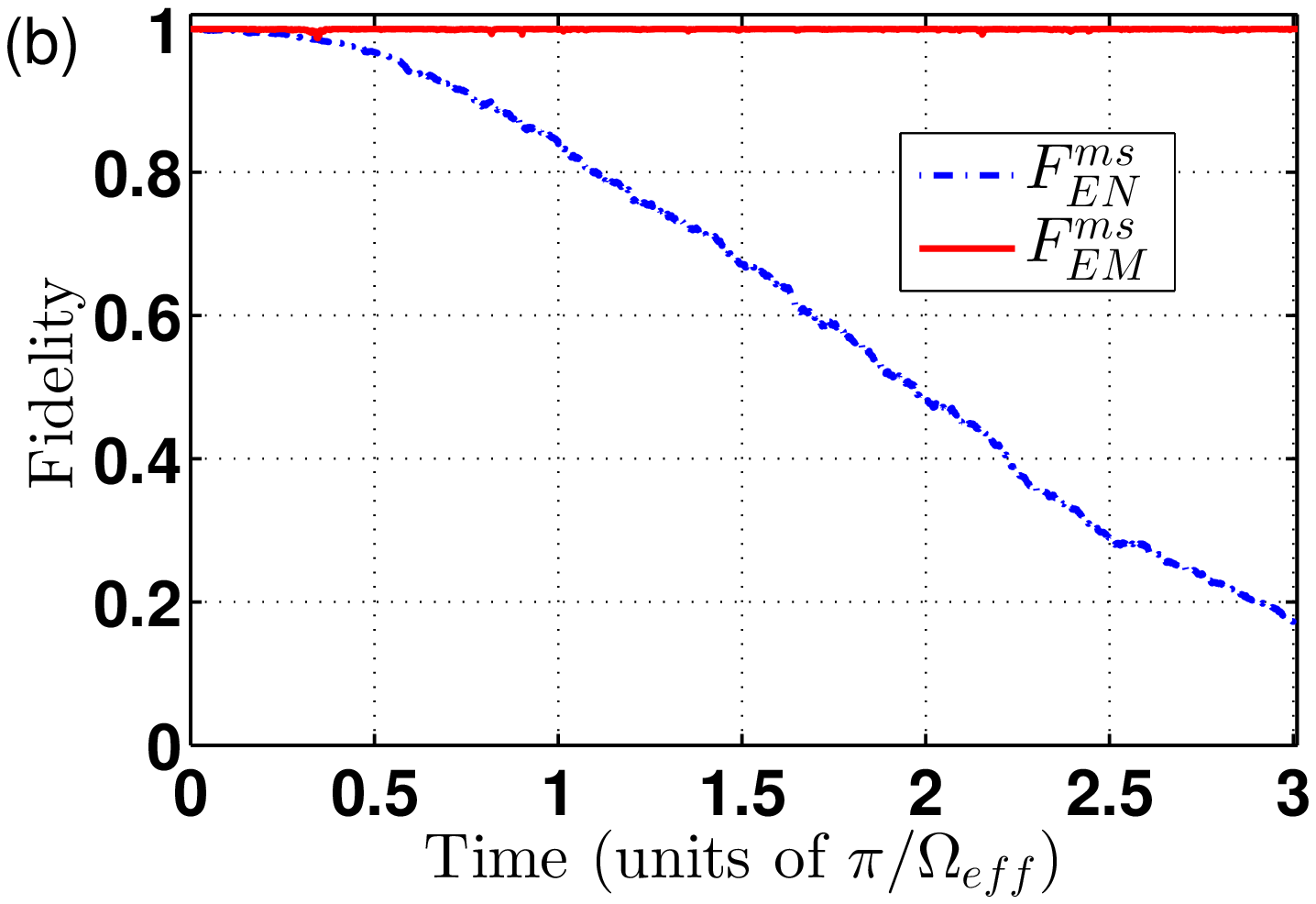}}
\caption{ The time evolution of fidelities when considering time-varying noises in the case of multi-level quantum system. The parameters for the simulations are: $\Omega=2\pi\times15$ MHz, $\Omega_1=\Omega_2=\Omega$, $V=\Omega$, $\Delta=2\pi\times$740 MHz, $\mu=0.05\Omega$, $\sigma=0.1\Omega$, $K=3500$, $\tau=1$ $ns$, and $\{p_{1, 2, 3, 4}, p_{5, 6, 7, 8}, p_9\}$=$\{1/18, 3/18, 1/9\}$.}\label{RydFed}
\end{figure}

At large intermediate-state detuning, $\Delta\gg\Omega_j$ ($j=1, 2$), the single atom state $|e\rangle$ is scarcely populated, and it can be adiabatically eliminated in the subspace of states $|g\rangle$ and $|r\rangle$ \cite{NVV1997,YiC2016}.
Then, after moving $\hat{H}_I$ to the rotating frame with respect to $\hat{U}=e^{-i\Delta t\sum_{j=1}^2|e\rangle_{j}\langle e|}$, the effective Hamiltonian of the system is
\begin{eqnarray}\label{}
\hat{H}_{eff}=\sqrt{2}\Omega_{eff}(|gg\rangle\langle T| + |rr\rangle\langle T| + H.c.) + V |rr\rangle\langle rr|,\nonumber\\
\end{eqnarray}
where $\Omega_{eff}=\Omega_1\Omega_2/\Delta$ and $|T\rangle=(|gr\rangle+|rg\rangle)/\sqrt{2}$.
Furthermore, we move the effective Hamiltonian $\hat{H}_{eff}$ to the rotating frame with respect to $\hat{U}=e^{-iVt|rr\rangle\langle rr|}$, then
\begin{eqnarray}\label{}
\hat{H}_{eff}=\sqrt{2}\Omega_{eff}(|gg\rangle\langle T| + |rr\rangle\langle T|e^{iVt} + H.c.).
\end{eqnarray}
If the parameters satisfy $V\gg\sqrt{2}\Omega_{eff}$, the fast oscillating term can be ignored safely.
That is, the doubly excited Rydberg state $|rr\rangle$ cannot be pumped from $|T\rangle$, which is known as the Rydberg blockade \cite{DTo2004,MSa2005,XQS2017,SLS2017}.
Clearly, the population of the system then would oscillate between states $|gg\rangle$ and $|T\rangle$ if the initial state is $|gg\rangle$ (or $|T\rangle$).

In order to illustrate the capability of the measurement feedback scheme, we choose the undisturbed Hamiltonian  $\hat{H}_0=\hat{H}_I$ (Eq. (\ref{undistH})) and consider the situation that there exists time-varying noises in the laser fields $\Omega_1$ and $\Omega_2$.
Thus the $\hat{\mathcal{H}}_j$ in the noise Hamiltonian Eq. (\ref{NoisH}) are given as
\begin{eqnarray}\label{RydNH}
&&\hat{\mathcal{H}}_1= |e\rangle_1\langle g|\otimes\hat{\mathcal{I}}_2 +\hat{\mathcal{I}}_1\otimes|e\rangle_2\langle g|+H.c. \cr
&&\hat{\mathcal{H}}_2= |r\rangle_1\langle e|\otimes\hat{\mathcal{I}}_2 +\hat{\mathcal{I}}_1\otimes|r\rangle_2\langle e|+H.c..
\end{eqnarray}
We use the fidelity $F^{ms}_{TN}=|\langle T| \psi^N\rangle|^2$ of the entangled state $|T\rangle$ to represent the noise-disturbed evolution.
For contrast, we also use the fidelity $F^{ms}_{TE}=|\langle T|\psi^E\rangle|^2$ to denote the evolution of the noiseless system.
The evolution of both fidelities $F^{ms}_{TN}$ and $F^{ms}_{TE}$ are displayed in Fig. \ref{RydFed}(a) with the initial state $|\psi(0)\rangle=|gg\rangle$.
It is clearly shown that the initial state $|gg\rangle$ can be pumped to the entangled state $|T\rangle$ periodically (red solid line) in the absence of noises.
However, the original periodical evolution of the fidelity is destroyed when the system suffering from time-varying noises (green dash-dot line).
That is, it appears a random phase shift and amplitude fluctuation during evolution process.
As a result, the system cannot reach the entangled state $|T\rangle$ perfectly.
To show the influence of the noises on the fidelity more clear, we also quantify the resemblance between the exact evolution $|\psi^E\rangle$ and the noises-disturbed evolution $|\psi^N\rangle$ by the fidelity $F^{ms}_{EN}=|\langle\psi^E|\psi^N\rangle|^2$, as is shown by the blue dash-dot line in Fig. \ref{RydFed} (b).
One can note that the value of the fidelity $F^{ms}_{EN}$ gradually decreases, indicating the detrimental influence of the time-varying noises on the resemblance between the exact and actual dynamical trajectories.

In the following, we display the performance of the measurement-feedback scheme for eliminating the influence of the time-varying noises.
For such a composite system, the Kraus operator $\hat{M}_n^{(k)}$ as given by  Eq. (\ref{Effects}) can be constructed in the Hilbert space $\{|gg\rangle,|ge\rangle,|eg\rangle,|ee\rangle,|gr\rangle,|er\rangle,|rg\rangle,|re\rangle,|rr\rangle\}$.
If we label the above corresponding basis states as $\{|1\rangle,|2\rangle,|3\rangle,|4\rangle,|5\rangle,|6\rangle,|7\rangle,|8\rangle,|9\rangle\}$, the Kraus operator can also be formally represented by Eq. (\ref{Effects}), and the measurement result $n$ is the atomic state of the composite system.
For instance, the measurement result $``n=1"$ represents the two atoms both in the ground state $|g\rangle$, i.e. $|gg\rangle$.
As a contrast, we also display the result of the elimination of the time-varying noises in Fig. \ref{RydFed}, by defining the fidelity $F^{ms}_{TM}=|\langle T|\psi^N_M\rangle|^2$, where $|\psi^N_M\rangle$ is the state under measurement-feedback operations and is given by Eq. (\ref{NME}).
By comparing the evolution of the two fidelities $F^{ms}_{TE}$ and $F^{ms}_{TM}$ in Fig. \ref{RydFed}(a), we can see that they are coincidence with each other perfectly, including the oscillation period and the amplitude.
This point can also be confirmed by the fidelity $F^{ms}_{EM}=|\langle\psi^E|\psi^M_N\rangle|^2$ (the red solid line in Fig. \ref{RydFed}(b)), which shows the resemblance between the exact evolution $|\psi^E\rangle$ and the measurement-feedback operated evolution $|\psi^M_N\rangle$.
The numerical results demonstrate that the fidelity $F^{ms}_{EM}$ keeps higher than $0.99$ all the time, indicating the good performance of the measurement-feedback scheme in the multi-level system.

%%%%%%%
\section{Discussions}\label{discussions}

\subsection{Influence of noise strength}\label{INS}

In the above applications, we only consider the case of static noises with $\Omega_\varepsilon=0.05\Omega_L$ for the two-level system, in which the high fidelity qubit dynamical trajectory is obtained.
In realistic scenarios, the static noises are always unknown and wide-range.
Hence, thoroughly investigating the influence of a wide range of unknown static noises on qubit dynamics is highly desirable.
For simplicity, we consider a qubit system whose undisturbed Hamiltonian $\hat{H}_0$ and the static noise Hamiltonian $\hat{H}_N$ has the following form,
\begin{eqnarray}\label{HamiltQ}
\hat{H}_0 &=&\frac{\Omega_0}{2}\hat{\sigma}_x+\frac{\beta_0}{2}\hat{\sigma}_z, \cr
\hat{H}_N &=& \frac{\delta \epsilon}{2}\hat{\sigma}_x+\frac{\delta \beta}{2}\hat{\sigma}_z,
\end{eqnarray}
where $\Omega_0$ is the strength of driving field and $\beta_0$ is the energy splitting.
The constant $\delta \epsilon$ and $\delta \beta$ are unknown stochastic noises that are independent of each other.

\begin{figure}[htbp]
\centering
\scalebox{0.4}
{\includegraphics{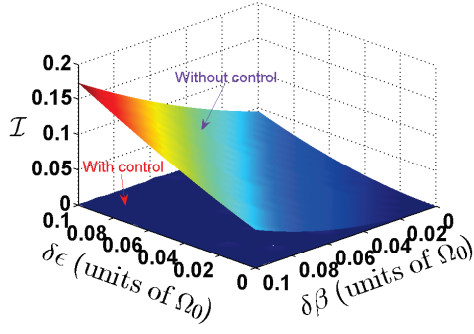}}
\caption{The infidelity $\mathcal{I}$ as a function of variations $\delta\beta$ and $\delta\epsilon$ with and without measurement-feedback operations. Parameters for the simulation are: $t_f=4\pi/\sqrt{\Omega^2_0+\beta^2_0}$, $\Omega_0=\beta_0$, $K=100$, $\tau=t_f/K$ and $p_0=0.25$.}\label{FError}
\end{figure}

\begin{figure}[b]
\centering
\scalebox{0.4}{\includegraphics{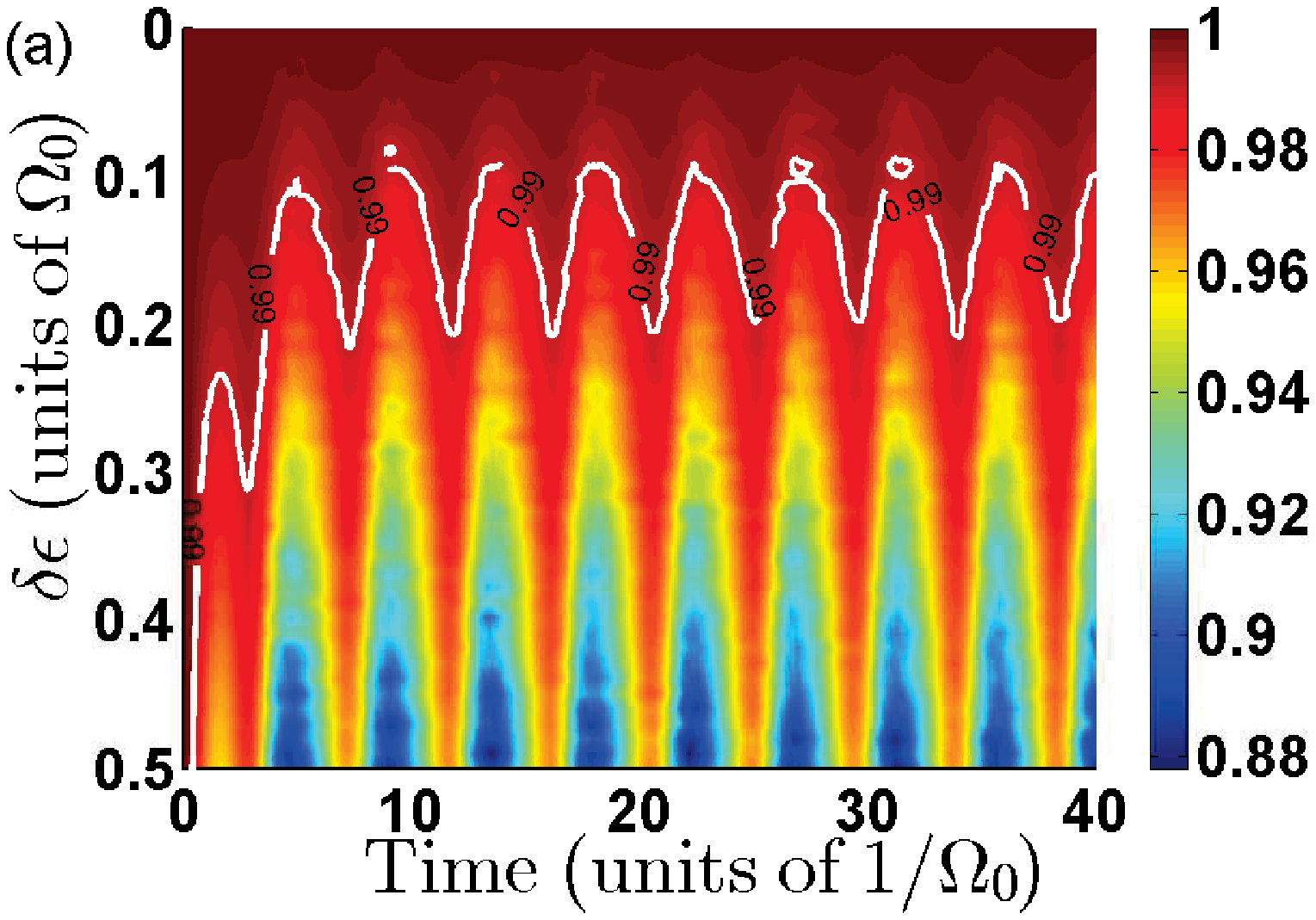}}
\scalebox{0.4}{\includegraphics{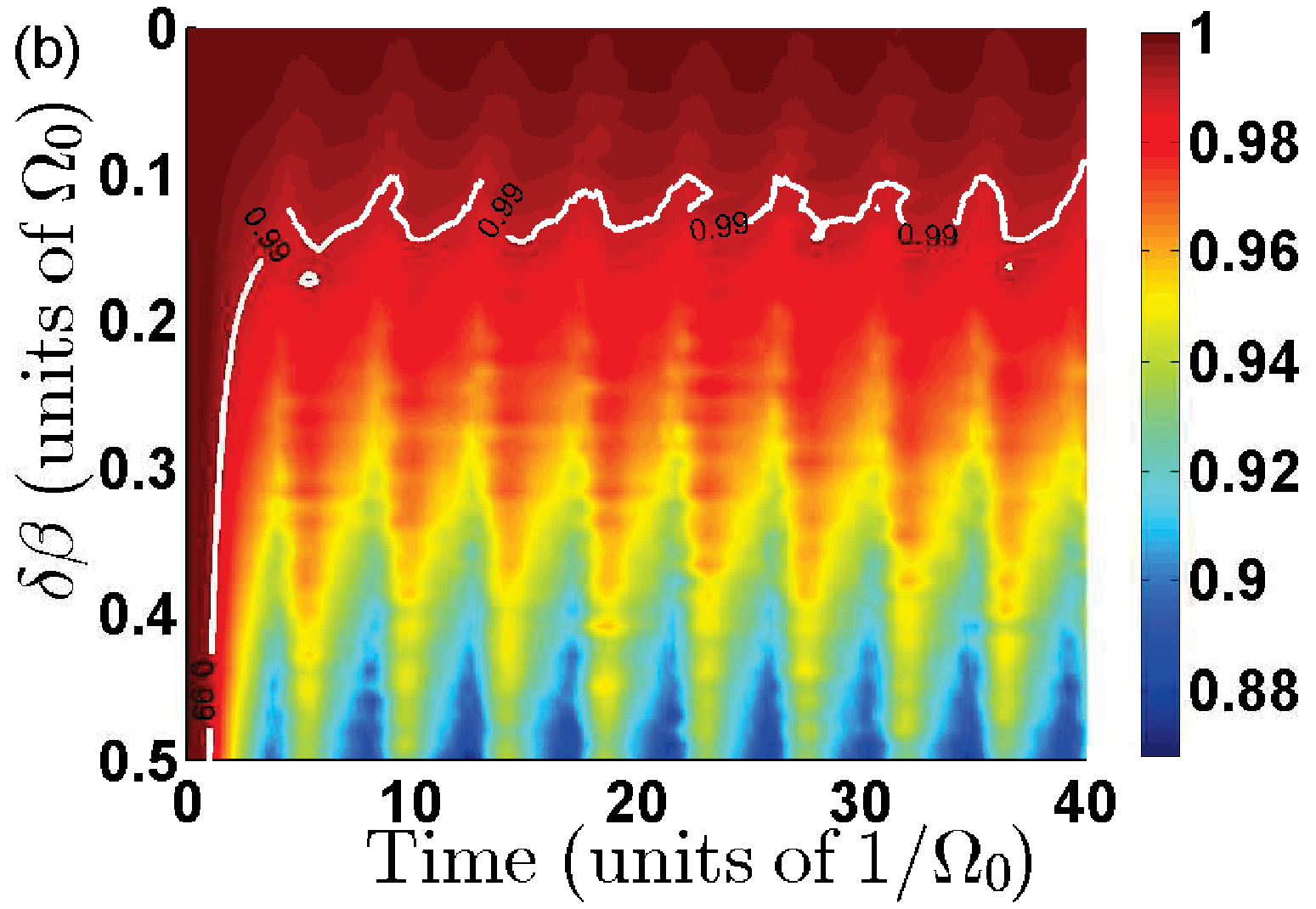}}
\caption{(a) The average fidelity $\overline{F^{qs}_{EM}}$ as a function of time and the stochastic noise $\delta\epsilon$ with $\delta\beta=0.05\Omega_0$. (b) The average fidelity $\overline{F^{qs}_{EM}}$ as a function of time and the stochastic noise $\delta\beta$ with $\delta\epsilon=0.05\Omega_0$. Other parameters for the simulation are: $T_L=2\pi/\sqrt{\Omega^2_0+\beta^2_0}$, $\Omega_0=\beta_0$, $K=500$, $\tau=T_L/50$ and $p_0=0.35$.}\label{NoiStre}
\end{figure}

In Fig. \ref{FError}, we plot the infidelity $\mathcal{I}=1-|\langle\psi^E(t_K)|\psi^N_{M}(t_K)\rangle|^2$ as a function of variations $\delta\beta$ and $\delta\epsilon$ with and without measurement-feedback operations, where $|\psi^E(t_K)\rangle$ is the state of the noiseless system at time $t_K$.
In the top surface of Fig. \ref{FError}, the fidelity error gradually increases with the increase of the value of $\delta\beta$ and $\delta\epsilon$ in the absence of measurement-feedback operations.
After the measurement-feedback operations are employed, the influence on $\mathcal{I}$ caused by stochastic noises $\delta\beta$ and $\delta\epsilon$ is almost eliminated (see from the bottom surface of Fig. \ref{FError}).
For instance, $\mathcal{I}\sim0.1727$ when $\{\delta\beta, \delta\epsilon\}\rightarrow0.1$ without control.
While it can be reduced to $\min\{\mathcal{I}\}\sim10^{-4}$ when measurement-feedback operations are imposed.
The almost vanishing value of $\mathcal{I}$ indicates that the high-fidelity qubit state can be obtained for a wide range of unknown stochastic noises at the specific time $t_K$ by using the measurement-feedback operations.
Actually, one cares more about the influence of the noise strength on qubit dynamical trajectory than a specific time.
Therefore, we show contour plots of the average fidelity $\overline{F^{qs}_{EM}}= \overline{|\langle\psi^E|\psi^N_M\rangle|^2}$ (averaged over 1000 runs) as a function of time and the stochastic noise $\delta\epsilon$ in Fig. \ref{NoiStre}(a) and $\delta\beta$ in Fig. \ref{NoiStre}(b).
We are fixing the initial state of the qubit system in $|\psi(0)\rangle=|+\rangle$ and $p_0=0.35$.
We see from Fig. \ref{NoiStre} that for small value of $\delta\epsilon$ and $\delta\beta$, such as $\delta\epsilon, \delta\beta<0.1\Omega_0$, the high-fidelity ($\overline{F^{qs}_{EM}}>0.99$) qubit dynamical trajectory can be maintained.
By increasing the value of $\delta\epsilon$ and $\delta\beta$, the average fidelity $\overline{F^{qs}_{EM}}$ displays significant oscillation behavior.
The observations indicate that for the given measurement strength (quantified by $p_0$ and $\tau$), the measurement-feedback scheme works well for the noise strength below 10\% of the driving field $\Omega_0$.
When the noise strength is larger than $0.1\Omega_0$, the qubit dynamical trajectory is gradually dominated by the noises other than the measurement-feedback operations.

\begin{figure}[b]
\centering
\scalebox{0.45}{\includegraphics{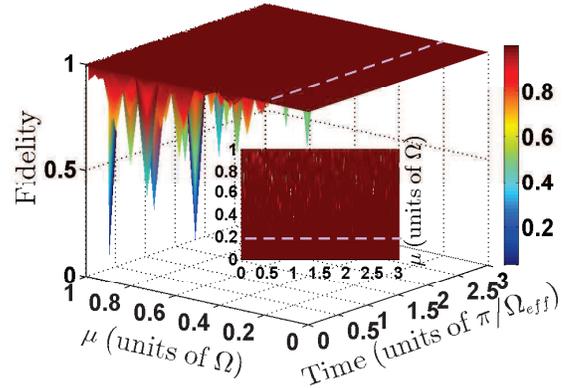}}
\caption{ The average fidelity $\overline{F^{mc}_{EM}}$ as a function of time and $\mu$ (the insert is the same figure from x-y view). Other parameters for the simulation are: The parameters for the simulations are: $\Omega=2\pi\times15$ MHz, $\Omega_1=\Omega_2=\Omega$, $V=\Omega$, $\Delta=2\pi\times$740 MHz, $\sigma=0.01\Omega$, $K=3500$, $\tau=1$ $ns$, and $\{p_{1, 2, 3, 4}, p_{5, 6, 7, 8}, p_9\}$=$\{1/18, 3/18, 1/9\}$.}\label{MulMu}
\end{figure}

In the above discussion, we consider the influence of noise strength of static noises on the qubit system dynamical trajectory.
It wonders how the strength of time-varying noises deteriorate the dynamical trajectory of quantum system under measurement-feedback operations.
Next, we are going to briefly discuss this point by taking the multi-level system as the example.
Here, we characterize the strength of the time-varying noises by the mean value $\mu$ and standard deviation $\sigma$.
The noise strength is large when increasing the value of $\mu$ and $\sigma$.
To quantitatively show the influence of the noise strength on the dynamical trajectory of the multi-level system, we plot the average fidelity
$\overline{F^{ms}_{EM}}= \overline{|\langle\psi^E|\psi^N_M\rangle|^2}$ (averaged over 1000 runs) as a function of time and $\mu$ in Fig. \ref{MulMu}, which demonstrates that the fidelity $\overline{F^{ms}_{EM}}$ can sustain high value ($>0.99$) when $\mu<0.2\Omega$.
When further increasing the noise strength, the fidelity $\overline{F^{ms}_{EM}}$ displays random fluctuations rather than oscillation behaviors.
Such an observation indicate that the measurement-feedback scheme can be used to eliminate the time-varying noises for a moderate range of noise strength at the given measurement strength.

\subsection{Influence of measurement strength}\label{}
We can see from Sec. \ref{INS}, for a given measurement strength, the measurement-feedback scheme does not perform very well when there exists strong background noises.
It is curiosity that whether we can improve the fidelity by adjusting the measurement strength.
In this subsection, we study the influence of the measurement strength on the system dynamical trajectory.

\begin{figure}[htbp]
\centering
\scalebox{0.45}{\includegraphics{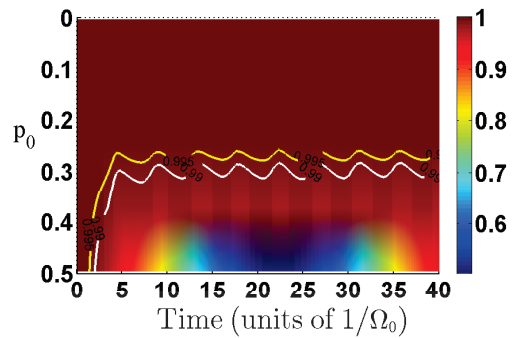}}
\caption{ The average fidelity $\overline{F^{qs}_{EM}}$ (averaged over 1000 runs) as a function of time and $p_0$. The parameters for the simulation are: $T_L=2\pi/\sqrt{\Omega^2_0+\beta^2_0}$, $\Omega_0=\beta_0$, $\delta\epsilon=\delta\beta=0.1\Omega_0$, $K=500$, and $\tau=T_L/50$.}\label{MeaStren}
\end{figure}

\begin{figure}[htbp]
\centering
\scalebox{0.48}{\includegraphics{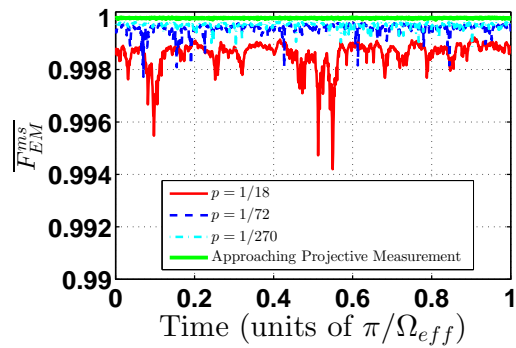}}
\caption{ The average fidelity $\overline{F^{ms}_{EM}}$ (averaged over 100 runs) as a function of time under different measurement strength. The parameters for the simulation are: $\Omega=2\pi\times15$ MHz, $\Omega_1=\Omega_2=\Omega$, $V=\Omega$, $\Delta=2\pi\times$740 MHz, $\mu=0.1$, $\sigma=0.1$, $\tau=1$ ns, $K=1167$. }\label{RydStre}
\end{figure}

Again, consider the qubit system suffering from static noises with the undisturbed Hamiltonian $\hat{H}_0$ and the static noise Hamiltonian $\hat{H}_N$ given by Eq. (\ref{HamiltQ}).
Notice that the measurement strength of the qubit system is stronger when the value of $p_0$ more approaching to zero ($\Delta p\rightarrow1$) for a fixed $\tau$.
Figure \ref{MeaStren} shows the average fidelity $\overline{F^{qs}_{EM}}=\overline{|\langle\psi^E|\psi^M_N\rangle|^2}$ as a function of time and $p_0$.
The results shows that there are large regions where the high-fidelity qubit dynamics can be achieved, such as $p_0<0.25$ ensures $\overline{F^{qs}_{EM}}>0.995$.
With increasing the value of $p_0$ in the region $0.25<p_0<0.35$, the average fidelity $\overline{F^{qs}_{EM}}$ exhibits small oscillations.
When further increasing the value of $p_0$ from 0.35 to 0.5, we can never maintain the high-fidelity.
Thus for the given static noises, $\delta\epsilon=\delta\beta=0.1\Omega_0$, the measurement-feedback operations can maintain high-fidelity qubit dynamics with moderate measurement strength ($p_0<0.35$).
The larger the measurement strength is, the higher the average fidelity will be.
This means that the measurement accuracy (or the amount of information extracted from the system) is closely related to the degree of precise for the dynamical trajectory of quantum system.
Such declaration is also correct for the multi-level system case within time-varying noises, which is illustrated in Fig. \ref{RydStre} by the average fidelity $\overline{F^{ms}_{EM}}=\overline{|\langle\psi^E|\psi^M_N\rangle|^2}$.
Notice that the measurement strength increases with the value of $p_j$ further deviates away from $1/9$ (i.e., the measurement strength is stronger when the value of $|p_j-1/9|$ is larger).
For simplicity, we suppose $p_1=p_2=p_3=p_4\equiv p$, $p_5=p_6=p_7=p_8\equiv1-p-p_9$ and $p_9=1/9$ in the simulation.
One can observe that the larger the value of $|p-1/9|$ is, the higher the averaged fidelity $\overline{F^{ms}_{EM}}$ will be.
When the measurement strength is approaching the projective measurement, the unite averaged fidelity is achieved.
Thus, the degree of precise for the dynamical trajectory of quantum system can be controlled by adjusting the measurement strength.

%%%%%%%
\subsection{Experimental feasibility}\label{}

In this subsection, we mainly discuss the experimental feasibility of the measurement-feedback scheme in eliminating the influence of background noises.
To implement unsharp measurements on a target system, generally, an auxiliary system is required \cite{SCo2018}.
When the target system interacts with the auxiliary system, one can yield universal control on the target system by manipulating the auxiliary system \cite{SLl2001,SLl2004}.
To be specific, the unsharp measurements on the target system are obtained by performing projective measurements on the auxiliary system.
Particularly, the measurement strength can be adjusted by regulating the interaction strength between target system and auxiliary system.
%The stronger the measurement strength is, the more information about the target system we obtain.
Experimentally, unsharp measurements on nuclear spin system have been achieved via single-shot readout with its measurement strength mapped on electron spin (ancilla) rotation angle \cite{JTM2018}.
On the other hand, measurement-based feedback operations can be realized by performing external control on the target or auxiliary system. For instance, an external field is added to an auxiliary system to achieve feedback control in generating and stabilizing Bell states \cite{LMa2018}, and cooling a mechanical resonator to its quantum ground state \cite{MBo2018}.
Alternatively, the unitary feedback $\hat{U}^{(k)}_n$ can be implemented by the evolution operator $\hat{\mathcal{U}}^{(k)}_n$ of an extra feedback Hamiltonian $\hat{H}^{(k)}_n$, i.e.,
\begin{eqnarray}\label{FedU}
\hat{U}^{(k)}_n=\hat{\mathcal{U}}^{(k)}_n=\exp{(-i\hat{H}^{(k)}_n{t_F})},
\end{eqnarray}
where ${t_F}$ represents feedback time and it is assumed to be very short (instantaneous feedback).
The experimental realization of the unitary feedback operation after a measurement has also been implemented \cite{GGG2010}.

\begin{figure}[htbp]
\centering
\scalebox{0.45}{\includegraphics{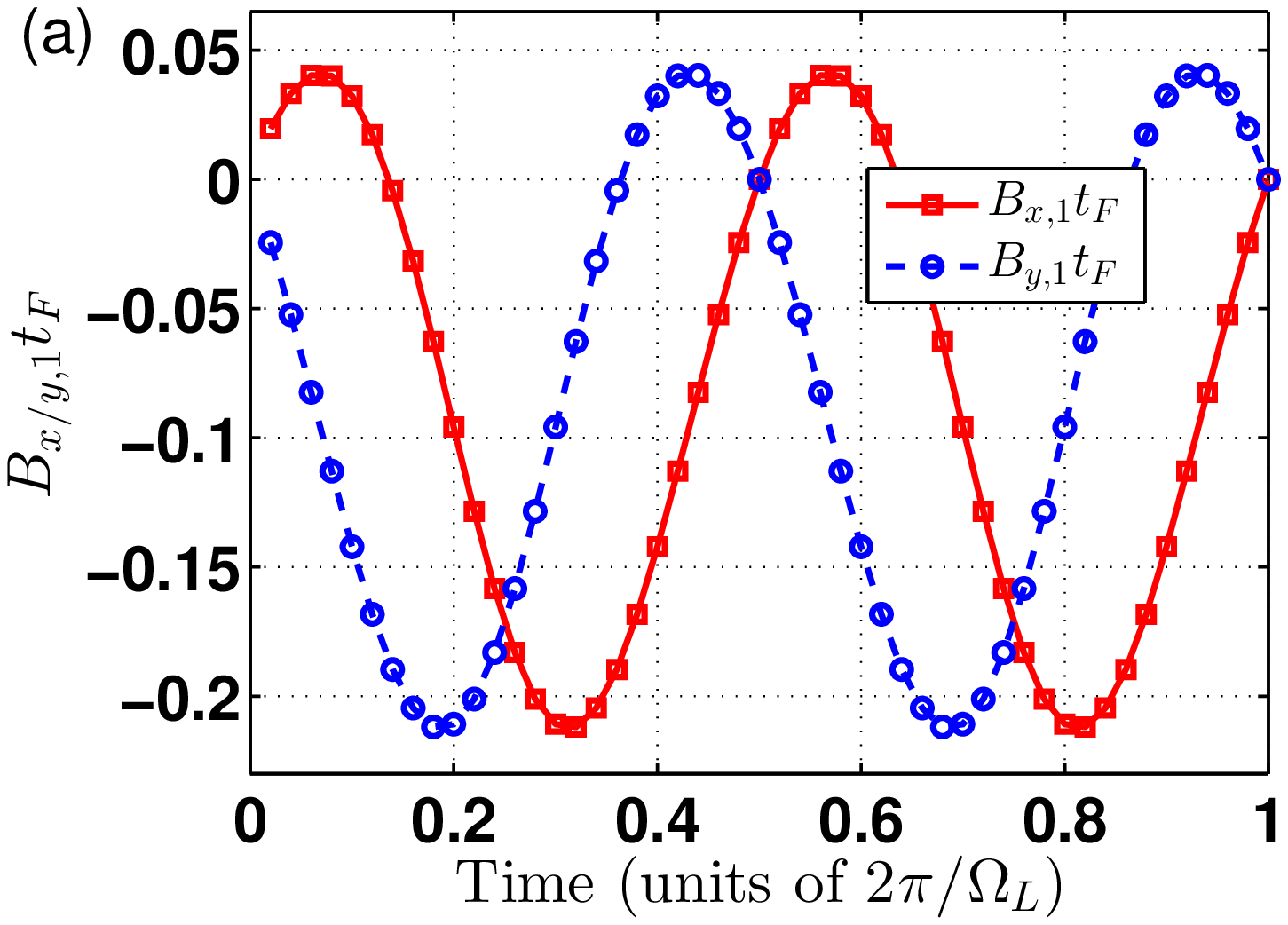}}
\scalebox{0.45}{\includegraphics{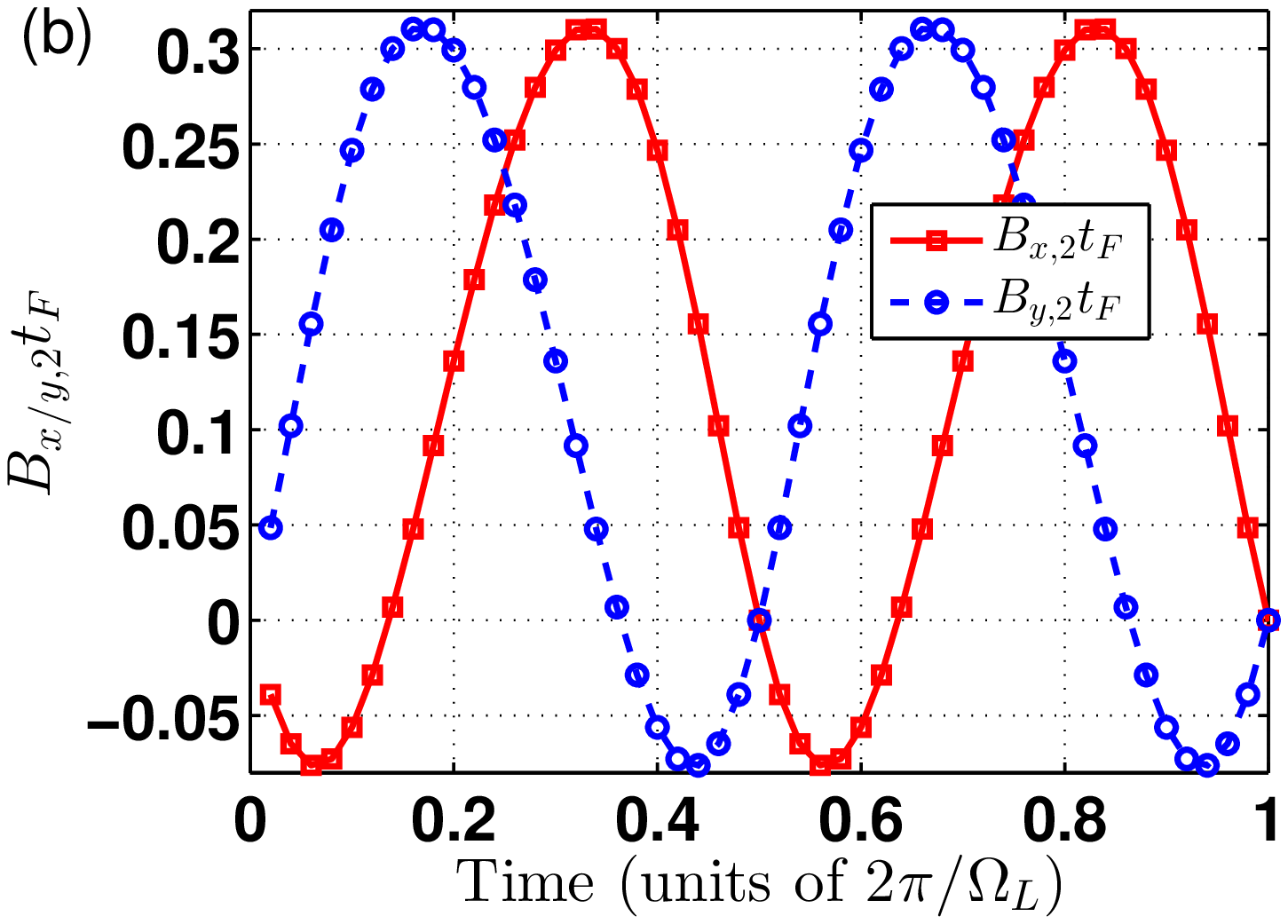}}
\caption{ Specific values of $B_{x/y,1}t_F$ (a), and $B_{x/y,2}t_F$ (b) versus time $k\tau$ for implementing unitary feedback operation when eliminating the influence of static noises in spin system. The parameters are: $p_0=0.2$, $K=50$, and the remaining parameters are the same of Fig. \ref{FeildB}.}\label{FeeH}
\end{figure}

Next, we take the system of $^{13}$C nuclear spin in diamond as an example to show the detailed procedures of measurement-feedback scheme in experiment.
Here, the unsharp measurements of the nuclear spin are realized by coupling it to an electron spin (e.g., two energy levels of a nitrogen-vacancy electron spin) and then performing projective measurements on the electron spin.
%For simplicity, we consider the nuclear spin under an external magnetic field along $z$-axis only.
The Hamiltonian of the whole system, including the electron spin ($S = 1/2$) and the nuclear spin ($I = 1/2$), reads \cite{GQL2017}
\begin{eqnarray}\label{}
\hat{H}=g\hat{S}_z\hat{I}_z + \Omega_L\hat{I}_z,
\end{eqnarray}
where $\hat{S}_z$ ($\hat{I}_z$) is the electron (nuclear) spin operator with eigenstates $|\pm\rangle_e$ ($|\pm\rangle_n$), $g$ is the coupling strength between the nuclear spin and the electron spin, and $\Omega_L$ is the Larmor frequency of the nuclear spin.
The evolution operator of the whole system after applying the Ramsey sequence is \cite{PCM2012}
\begin{eqnarray}\label{}
\hat{\mathbf{U}}(t) = \hat{R}^x_e(\frac{\pi}{2})[\hat{U}^{(+)}_n(t)|+\rangle_e\langle +| + \hat{U}^{(-)}_n(t)|-\rangle_e\langle -|]\hat{R}^y_e(\frac{\pi}{2}),\nonumber\\
\end{eqnarray}
where $\hat{R}^h_e(\frac{\pi}{2})=e^{-i\hat{S}_h\pi/2}$ ($h=x, y$) denotes the $\pi/2$ pulse for the electron spin along $h$-axis, and $\hat{U}^{(\pm)}_n(t)=e^{-i(\Omega_L\pm g/2)\hat{I}_zt}$ is the evolution operator of the nuclear spin conditioned on the electron spin state.
Suppose the initial state of the whole system is $|\psi(0)\rangle=|+\rangle_e\otimes|\psi(0)\rangle_n$ with $|\psi(0)\rangle_n=a_0|+\rangle+b_0|-\rangle$  ($|a_0|^2+|b_0|^2=1$) denoting the initial state of the nuclear spin, then performing projective measurements on the electron spin with the projective measurement operator $\hat{M}^{(\alpha)}_e=(\hat{\mathbb{I}}+2\alpha \hat{S}_z)/2$ ($\alpha=\pm 1$) is equivalent to an unsharp measurement on the nuclear spin, i.e.,
\begin{eqnarray}\label{}
\hat{M}^{(\alpha)}_n\hat{\rho}(0)(\hat{M}^{(\alpha)}_n)^{\dagger}
=\textrm{Tr}_e[\hat{M}^{(\alpha)}_e\hat{\mathbf{U}}(t)\hat{\rho}(0)\hat{\mathbf{U}}(t)(\hat{M}^{(\alpha)}_e)^{\dagger}], \nonumber\\
\end{eqnarray}
where $\hat{\rho}(0)=|\psi(0)\rangle\langle\psi(0)|$ and
\begin{eqnarray}\label{}
\hat{M}^{(\alpha)}_n &=& \frac{1}{2}[\hat{U}^{(+)}_n(t) -i\alpha \hat{U}^{(-)}_n(t)]  \cr
&=& e^{-i(\Omega_L\hat{I}_zt+\alpha\frac{\pi}{4})}\frac{[\cos(\vartheta)\hat{\mathbb{I}}+2\alpha\sin(\vartheta)\hat{I}_z]}{\sqrt{2}},
\end{eqnarray}
with $\vartheta=gt/2$.
We can see that $\hat{M}^{(\alpha)}_n$ is equivalent to the measurement operator as that given in Eq. (\ref{MeOp}) with $p_0=\frac{1}{2}[\cos(\vartheta)-\sin(\vartheta)]^2$, excepting an additional phase factor $e^{-i(\Omega_L\hat{I}_zt+\alpha\frac{\pi}{4})}$ that has no effect on the probability distribution of the measurement results.
Note that the existence of background noises on the Larmor frequency only influence the additional phase factor, and thus has no effect on the probability distribution of the measurement results as well.
For instance, when $\Omega_L\rightarrow\Omega_L+\Omega_{\varepsilon}$, the additional phase factor becomes $e^{-i[(\Omega_L+\Omega_{\varepsilon})\hat{I}_zt+\alpha\frac{\pi}{4}]}$.
By repetitively applying the Ramsey sequence to the electron spin, we achieve a sequential unsharp measurements on the nuclear spin with the measurement strength depending on the time delay $t$ and the coupling strength $g$.
Therefore, unsharp measurements on the nuclear spin is achievable in the presence of the background noises.
Then, the residual crucial point of the scheme is the realization of unitary feedback operations.
In practice, it can be realized by utilizing simply-designed auxiliary feedback Hamiltonians $\hat{H}^{(k)}_n$, which are reversely solved from the unitary feedback $\hat{U}^{(k)}_n$ according to Eq. (\ref{FedU}).
For instance, in the elimination of the static noises with the parameters $p_0=0.2$ and $\tau=2\pi/(50\Omega_L)$, the concrete form of Hamiltonians $\hat{H}^{(k)}_n$ ($n=1, 2$ and $k=1, 2, \cdots, K$) in the basis $\{|+\rangle, |-\rangle\}$ is given by
\begin{eqnarray}\label{FedH}
\hat{H}^{(k)}_n=
\left
[
{\begin{array}
{*{20}{c}}
B^{(k)}_{z,n} &  B^{(k)}_{x,n} + iB^{(k)}_{y,n} \\
B^{(k)}_{x,n} - iB^{(k)}_{y,n}&  -B^{(k)}_{z,n}
\end{array}}
\right
],
\end{eqnarray}
where $B^{(k)}_{z,n}=0$ ($n=1, 2$), and the specific real values of $B^{(k)}_{x,n}{t_F}$ and $B^{(k)}_{y,n}{t_F}$ versus time $t_k$ are plotted in Fig. \ref{FeeH}.
In fact, the feedback Hamiltonian is easily implemented by adding extra magnetic fields to the nuclear spin along $h$-axis ($h=x, y, z$), where the corresponding amplitudes are $B^{(k)}_{h,n}/t_F$.
Thus the implementation of the measurement-feedback scheme for precise quantum system control can be available in practice.

\section{Conclusion}\label{conclusions}

In conclusion, we have proposed a scheme to precisely control system dynamics based on sequential unsharp measurements and unitary feedback operations.
The unsharp measurements are assumed to be the impulsive measurements and carried out periodically.
The feedback operation is well designed to recover the state of DTPMES back into its pre-measurement state.
For its applications, we first demonstrate to eliminate the influence of static noises in typical qubit quantum systems,
and then generalise to eliminate the influence of time-varying noises in multi-level systems.
Simulation results show that the scheme works well in eliminating different kinds of noises.
In particular, we can obtain high degree of precise for quantum system dynamics by increasing the measurement strength.
Additionally, there is no need to worry about the back-action effect of measurements, because it will be compensated by the tailored feedback operations.
Furthermore, with the well-designed unitary feedback operation, the scheme may be applied to different quantum systems for superposition state generation and quantum system dynamics elicitation, which are our further research topic.
Thus the scheme is powerful in quantum system control and is expected to have broad applications within quantum information processing.

%%%%%%%
\section*{Acknowledgments}
The authors wish to acknowledge helpful discussions with B. H. Huang, Y. H. Kang, and Z. L. Zhang.
This work was supported by the National Natural Science Foundation of China under Grants No. 11575045, No. 11747011, No. 11805036, and No. 11674060,
the Major State Basic Research Development Program of China under Grant No. 2012CB921601, the Fund of Fujian Education Department under Grant No. JAT170086, and the Natural Science Foundation of Fujian Province of China under grant No. 2018J01413.
%%%%%%

%%%%%%%

%%%%%%%
\end{document}